# MOBILE BASED SECURE DIGITAL WALLET FOR PEER TO PEER PAYMENT SYSTEM

Majid Taghiloo[1,] Mohammad Ali Agheli [2,] and Mohammad Reza Rezaeinezhad[3]

[1]Amnafzar Department of Pishgaman Kavir Yazd, Tehran, Iran
mtaghiloo@amnafzar.net
[2]Amnafzar Department of Pishgaman Kavir Yazd, Yazd, Iran
agheli@amnafzar.net
[3] Pishgaman Kavir Yazd, Yazd, Iran
rezaei@pishgaman.com

## ABSTRACT

E-commerce in today's conditions has the highest dependence on network infrastructure of banking. However, when the possibility of communicating with the Banking network is not provided, business activities will suffer. This paper proposes a new approach of digital wallet based on mobile devices without the need to exchange physical money or communicate with banking network. A digital wallet is a software component that allows a user to make an electronic payment in cash (such as a credit card or a digital coin), and hides the low-level details of executing the payment protocol that is used to make the payment. The main features of proposed architecture are secure awareness, fault tolerance, and infrastructure-less protocol.

## 1. INTRODUCTION

A digital wallet allows users to make electronic commercial transactions swiftly and securely. It functions much like a physical wallet. A digital wallet has both a software and information component. The software provides security and encryption for personal information and for the actual transaction. Typically, digital wallets are stored on the client-side and are easily compatible with most e-commerce transactions. A server-side digital wallet, known as thin wallet, is the one that an organization creates for you and maintains on its servers. The information component is basically a database of user inputted information. This information consists of your shipping address, billing address, and other information.

This concept provides a means by which customers may order products and services online without ever entering sensitive information and submitting it via wireless communication, where it is vulnerable to theft by hackers and other cyber-criminals.

The simplicity of financial transactions for every society is very important. In traditional methods, business will be done by exchanging physical money. Disadvantages of this method are quite evident. Besides transmission of diseases, the physical security challenges of this method are undeniable. Technology improvement and expansion of communications networks have been thriving e-commerce affairs.

In order to solve these kinds of problems, various banks have developed smart card and ATM system thay can withdraw money from business transactions. In this method, communication infrastructure is necessary for financial exchange. In many cases, because of communication failures, users notice the error "unable to connect to central server" on ATM display monitor. Communication infrastructure defects should not make challenge for customers and availability of banking service is a very important parameter, in this respect.

Hence, creating an independent method from infrastructure in order to exchange cash is very important. That way, the money of the customer should be kept virtually. A solution would be to replace the physical wallet with a digital wallet integrated into an existing mobile device like a cell phone. When the customer needs to perform financial transaction, the value of the stored virtual money should be updated. Digital money can be placed on hardware chip or stored as software data. Each of these methods has its own advantages and disadvantages. Creating independent chip or a dedicated embedded system for e-wallet will be costly. Also, its holding is very hard for customers.

In this paper, the software method based on mobile devices will be proposed to create e-wallet system. Communication media between mobile devices is wireless. Since there are a number of important challenges in the mentioned system, it must be managed especially by proposed protocol. Weather conditions can affect the strength of wireless signals, and it makes weakness on the robustness of wireless communications. Therefore, the protocol must consider the possibility of packet loss during the protocol packets exchange.

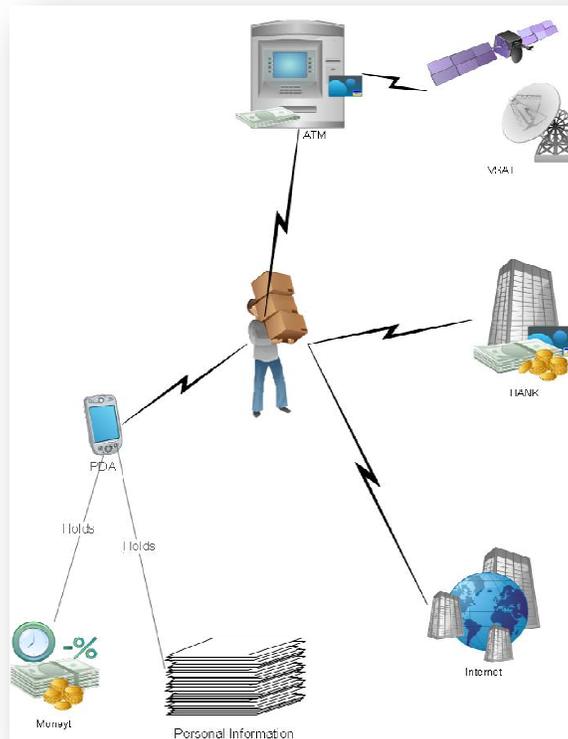

Figure 1. all possible facility for money exchange

Digital money is transferred in a distributed transaction manner. Here, distributed transaction is a transaction that updates data on two mobile based computer systems. Distributed transactions extend the benefits of transactions to applications that must update distributed data. Implementing robust distributed applications is difficult because these applications are subject to multiple failures, including failure of the seller node, the buyer node, and the network connection among them. This distributed transaction must be atomic. In other words, each transaction is said to be atomic if when one part of the transaction fails, the entire transaction fails and system state is left unchanged. Such transactions cannot be subdivided, and must be processed in its entirety or not at all.

Security is another challenge in order to maintain and manage the electronic money and some personal information of its owner. Moreover, security would be enhanced as all data on the digital wallet would be encrypted and back up options would make recovering from loss easier. So, the protocol should propose solution for all possible security threats in order to improving the reliability of the system.

However, the idea of digital wallet is not new. Indeed South Korea, America and Sweden have already rolled out digital-wallet based solutions [1] [2].

This paper describes the new approach for mobile payment system. The proposed solution can be implemented on Smart phone, and it does not require any additional connectivity or infrastructure beyond the cell phone of the participants, and was designed with usability and security in mind.

## 2. M-WALLET APPROACH

The following formatting rules must be followed strictly. This (.doc) document may be used as a template for papers prepared using Microsoft Word. Papers not conforming to these requirements may not be published in the conference proceedings.

Mobile systems are pervasively provided for society. This potential can be used to solve everyday problems. In this paper, a mobile device is used as a storing context for important personal data and digital money. One of the important reasons for selecting this context is widely use of mobile phones at community level and the second reason is sensitivity and protection of people regarding this tool.

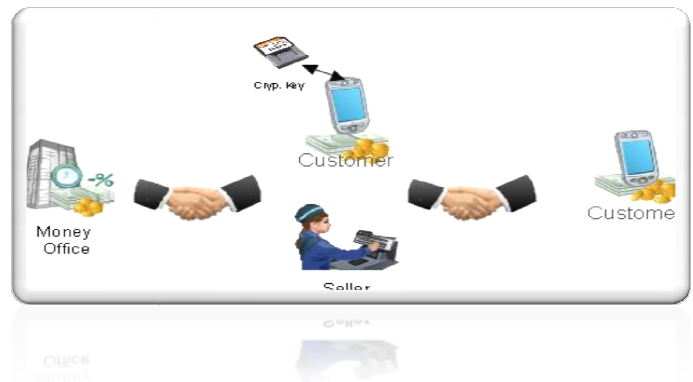

Figure 2. protocol entities

Entities involved in this system are: seller, buyer (customer), and money office.

## 2.1. Protocol Architecture

As shown in figure 2, mobile node will appear as buyer or seller and they can exchange digital money with each other. Also, central digital money charging office should be considered.

### 2.1.1. Digital Money Charge Scenario

Charging process will be done for equipping the mobile phone to digital money. In this process, secure information including digital money, information identification, public and private keys are stored in the individual phone. Therefore, there is no challenge for automatic key distribution, for describing security mechanism. Communication media is wireless link, so there are many security challenges that discussed in the following sections.

### 2.1.2. Financial Exchange Scenario

For digital money exchange, seller node based on using intelligent searching method, detects all surrounding buyer nodes. In order to do this task, it sends Look-up packet by one hop broadcasting message. Hence, all neighbour nodes will receive this message and then buyer nodes will response by accepting the message. Only neighbour node that is in buyer mode will response by accepting the message. It helps to save power of mobile node. Seller node will select the corresponding buyer by searching its unique ID. Then the cost of tools will be sent to buyer node. After receiving this message by buyer node, if the cost was correct and there is enough digital money, it will accept this request.

In this scenario there are many challenges, but for reducing the complexity, it will be described in separate sections of this paper. Security and distributed transaction are major challenges of this scenario. Figure 3 shows the sequence of messaging between buyer and seller nodes.

### 2.1.3. Money Exchange Scenario.

Process of digital money exchange between two separate nodes, introduces the concept of distributed transaction. This transaction is atomic. A transaction represents an atomic unit of work. Either all modifications within a transaction are performed, or none of the modifications are performed.

There are many researches about distributed transactions and the result is a two-phase commit protocol (2pc) and a three-phase commit protocol (3pc). Because of the high complexity of 3pc method, the 2pc solution has wide usage [8] [9] [10].

This paper proposes new efficient approach based on 2pc method, for managing distributed atomic transaction. In this method, all tasks distributed between only two nodes, Seller and Buyer, and also the Seller node plays as coordinator of distributed transaction. It makes the protocol very simple. Therefore, the 2pc protocol will be modified in the manner to do atomic transaction only between two nodes and it removes extra overhead of regular 2pc protocol.

The two phase commit protocol is a distributed algorithm which lets all sites in a distributed system agrees to commit a transaction. The protocol results in either all nodes committing the transaction or aborting, even in the case of node failures and message losses. The two phases of the algorithm are broken into the COMMIT-REQUEST phase, where the Seller node attempts to prepare Buyer node, and the COMMIT phase, where the Seller node completes the transactions at Buyer node.

The protocol works in the following manner: One node is designated as the coordinator, which is the master (Seller) node, and the other node is called Buyer node. One assumption of the protocol is stable storage at each node for storing the logs. Also, the protocol assumes that no node crashes forever, and eventually any two nodes can communicate with each other.

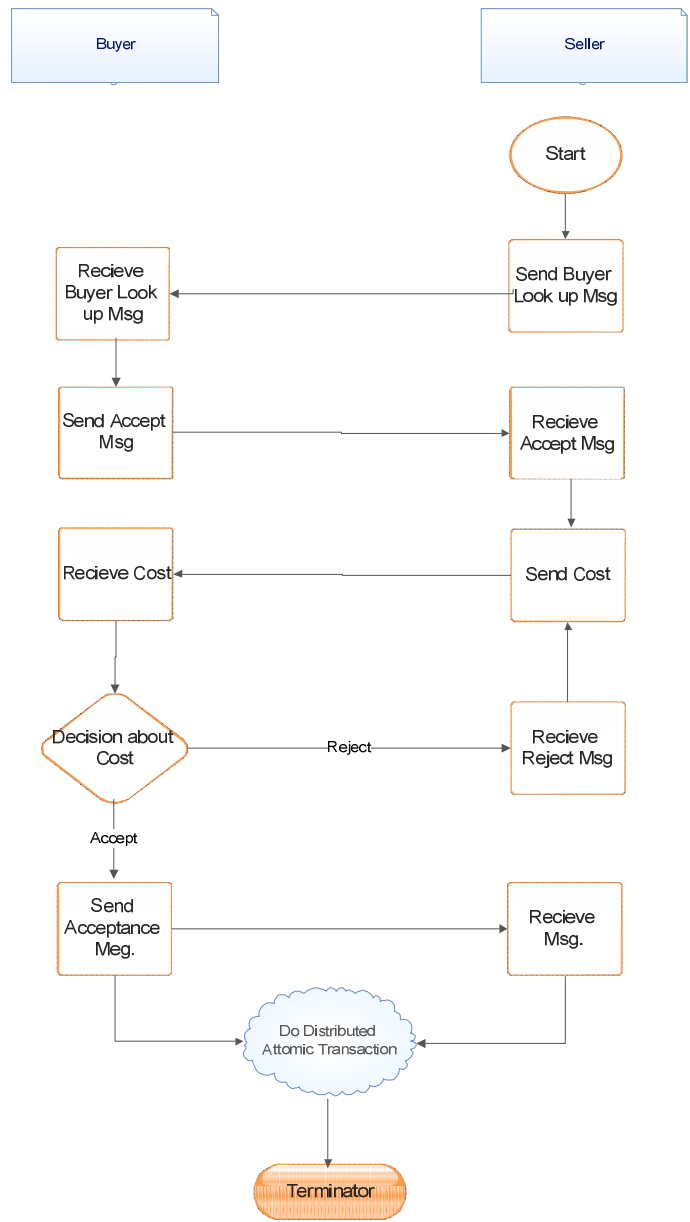

Figure 3. sequence of messaging for money exchanging

### 2.1.3.1. Basic Algorithm

During phase 1, initially the Seller node sends a query to *commit* message to Buyer node. Then it waits for Buyer node to report back with the agreement message. The Buyer node, if the transaction was successful, write an entry to the undo log and an entry to the redo log. Then the Buyer node replies with an *agree* message, or an *abort* if the transaction failed at Buyer node. During phase 2, if the Seller node receives an *agree* message from Buyer node, and then it writes a commit record into its log and sends a *commit* message to Buyer Node. If agreement message do not come back the Seller node sends an *abort* message. Next, the Seller node waits for the acknowledgement from the Buyer node. When acks are received from Buyer node the Seller node writes a complete record to its log. If the Buyer node receives a *commit* message, it releases all the locks and resources held during the transaction and send an acknowledgement to the Seller node. If the message is abort, then the Buyer node will undo the transaction with the undo log and releases the resources and locks held during the transaction. Then it sends an acknowledgement. Figures 4 and 5, demonstrate the *state diagram* of BUYER and SELLER nodes.

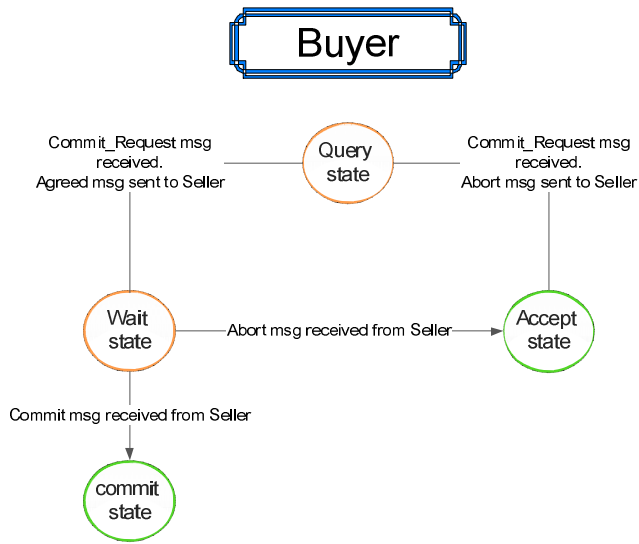

Figure 4. State diagram of Buyer Node

**2.1.3.2 The Detailed Commit Protocol**

At the SELLER node:

1. The SELLER node sends the message to the BUYER node. The SELLER node is now in the preparing transaction state.
2. Now the SELLER node waits for responses from the BUYER node. If the BUYER node responds ABORT then the transaction must be aborted, proceed to step 5. If the BUYER node responds AGREED then the transaction may be committed, and proceed to step 3. If after the number of period's expiration, the BUYER node does not respond, the SELLER node can either transmit ABORT messages to the BUYER node or transmit COMMIT-REQUEST messages to the BUYER node. In either case, the SELLER node will eventually go to state 3 or state 5.
3. Record in the logs a COMPLETE to indicate the transaction is now completing. Send COMMIT message to the BUYER node.
4. Wait for the BUYER node to respond. They must reply COMMIT. If after the number of period's expiration the BUYER node has not responded, retransmit the COMMIT message. Once the BUYER node has replied, erase all associated information from permanent memory. DONE.

Send the ABORT message to the BUYER node.

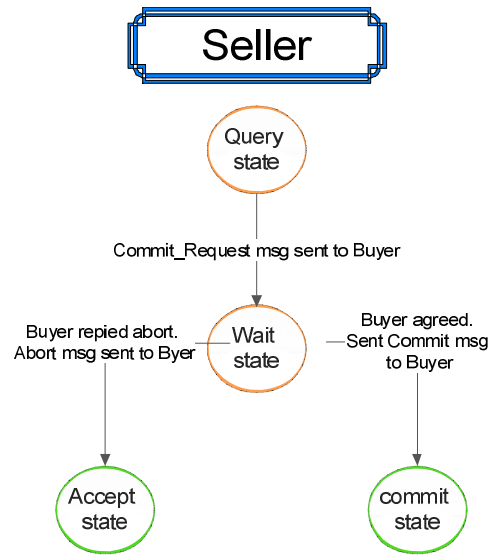

Figure 4. State diagram of Seller node

At the BUYER node:

1. If a COMMIT-REQUEST message is received for transaction T which is unknown at the BUYER node (never ran, eliminated by crash, etc), reply ABORT. Otherwise write the new state of the transaction to the UNDO and REDO log in permanent memory. This allows for the old state to be recovered (in event of later abort) or committed on demand regardless of crashes. The read locks of a transaction may be released at this time; however, the write locks are still maintained. Now send AGREED to the SELLER node.

2. If an ABORT message is received, then kill the transaction, which involves deleting the new state if the transaction from the REDO and UNDO log the new state of the transaction, and restore any state before the transaction occurs.

3. If a COMMIT message is received, then the transaction is either prepared for committal or already committed. If it is prepared, perform all the operations necessary to update the database and release the remaining locks the transaction possesses. If it is already committed, no further action is required. Respond COMMITED to the SELLER node.

## 2.5. Correctness

We assert that if the BUYER node completes the transaction, the SELLER node completes the transaction eventually. The proof for correctness proceeds somewhat informally as follows: If the BUYER node is completing a transaction, it is only because the SELLER node sent a COMMT message. This message is only sent when The SELLER node is in the commit phase, in which case the SELLER node itself has completed the transaction eventually. This means the SELLER node and the BUYER node have prepared the transaction, which implies any crash at this point will not harm the transaction data because it is in permanent memory. Once The SELLER node is completing, it is ensured that the BUYER node completes before the SELLER node's data is erased. Thus, crashes of the SELLER node do not interfere with the completion.

Therefore if the BUYER node completes, so does the SELLER node. The abort sequence can be argued in a similar manner. Hence, the atomicity of the transaction is guaranteed to fail or complete globally.

### 2.6. Security Mechanism

Increasingly, companies and individuals are using wireless technology for important communications they want to keep private, such as mobile e-commerce transactions, e-mail, and corporate data transmissions.

At the same time, as wireless platforms mature, grow in popularity, and store valuable information, hackers are stepping up their attacks to this new targets. This is particular problematic, because wireless devices, including smart cellular phone and personal digital assistants (PDAs) with Internet access, were not originally designed with security as top priority. Now, however, wireless security is becoming an important area of product research and development.

As in the wired world, wireless security boils down to protect information and prevent unauthorized system access. However, it is challenging to implement security in small-footprint devices with low processing power and small memory capabilities which use unreliable, low bandwidth wireless networks.

A key aspect of security for activities such as mobile e-commerce and mission-critical corporate communications is the ability to authenticate a message the sender's identity. Digital money and personal information of mobile device owner is very sensitive data. In order to apply authentication and confidentiality service for protecting this kind of sensitive data, the paper proposes new security scheme.

The proposed cryptographic system has three fundamental features: confidentiality, integrity, and non-repudiation. These features combine to provide authenticity [11]. Confidentiality means that the message contents remain private. The plaintext cannot be viewed by anyone who does not have the necessary keys, algorithms, and tools. Integrity refers to keeping a message unchanged. And non-repudiation means that a person cannot deny signing a particular message, which is especially important in the context of transaction.

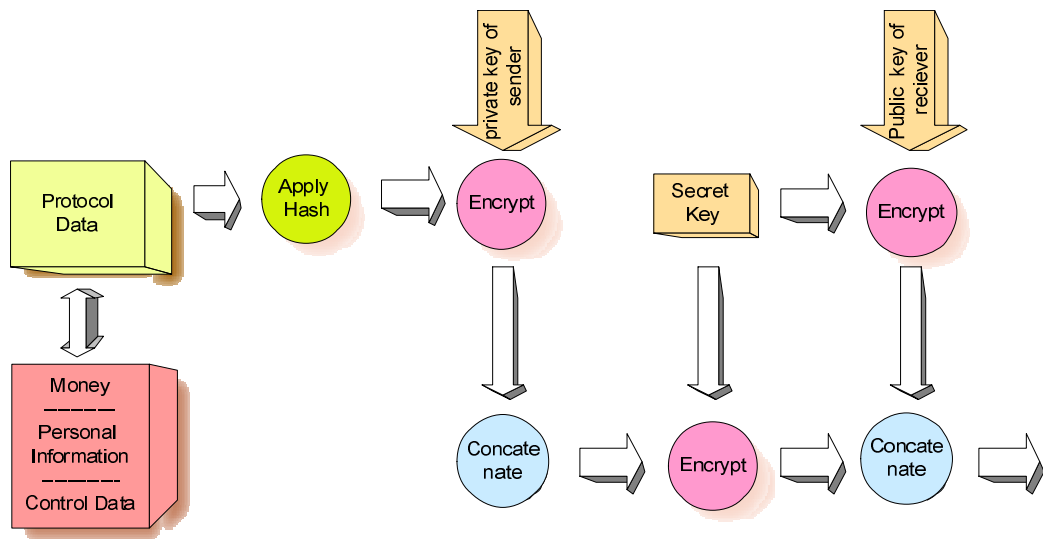

Figure 5. Applying security mechanism at sender

Security mechanism at sender node:

1. Apply hash function to the message
2. Encrypt the result of hash function with private key of sender
3. Concatenate the result of step 2 with original data
4. Encrypt result of step 3 with secret key
5. Encrypt secret key by public key of receiver
6. Concatenate step 4 and 5

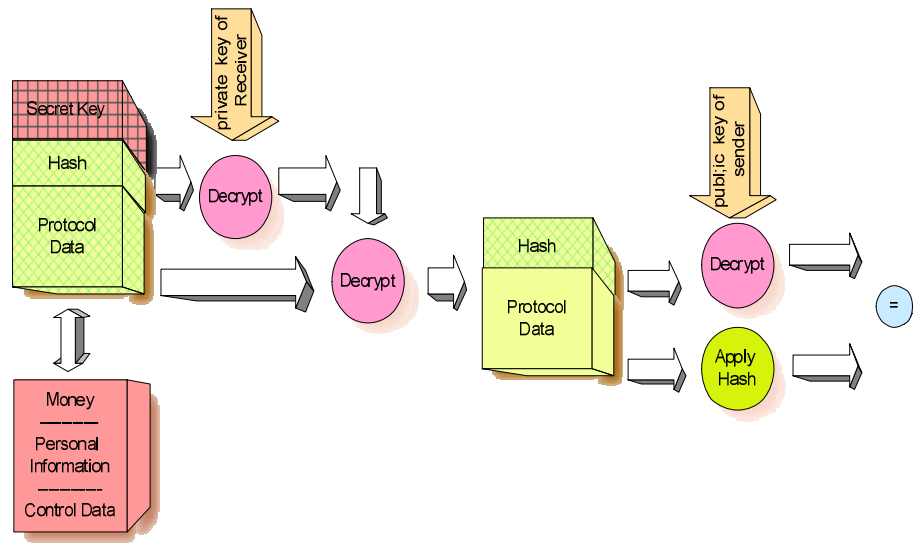

Figure 6. Applying security mechanism at receiver

Security mechanism at receiver node:

1. Extract secret key and decrypt it by receiver's private key
2. Decrypt message by result of step 1 (secret key)
3. Decrypt hash by sender's public key
4. Apply hash to the remained data
5. Compare the result of step 3 and step 4. If the result was equal then it means that data was send in secure manner and there is no threat in this respect. Otherwise, the received message is not valid.

## 3. CONCLUSION

In this paper, we presented the design of architecture of Mobile based Digital Wallet for peer to peer payment system. Proposed solution is encryption software that works like a physical wallet during electronic commerce transactions. It can hold a user's payment information, a digital certification to identify the user, and shipping information to speed transactions. The consumer benefits because his or her information is encrypted against piracy and because some wallets will automatically input shipping information at the merchant's node and will give the consumer the option of paying by digital cash or check.

In near future, we plan to implement this architecture in a real environment. We also plan to test the security scheme of proposed protocol.

## ACKNOWLEDGEMENTS

The authors would like to thank Pishgaman Kavir Yazd Corporation for financial support and also would like to thank Ms Hasti Tabrizi Nasab for valuable comments. Lastly, we offer our regards and blessing to all of those who supported us in any respect during the completion of the paper.

**Authors**

Majid Taghiloo is currently a researcher of Amnafzar Department of Pishgaman Kavir Yazd Co. He completed his degree in Information Technology Engineering, in 2008, at Amirkabir University of Technology. He did his BSc in Computer Software Engineering, in 2004 at Tehran University of IRAN. His areas of interests include Security, Ad Hoc Network, Information Technology and E-Commerce. He is now a lecturer of Islamic Azad University of Tehran, South Branch.

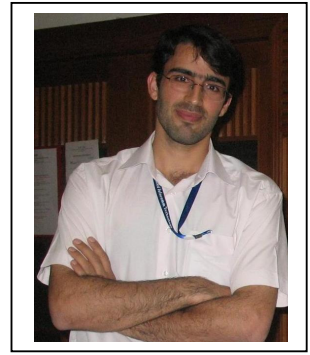

Mohammad Ali Agheli Hajiabadi is currently a Managing Director of Amnafzar department of Pishgaman Kavir Yazd Co. He completed his degree in Information Technology Engineering, in 2011, at Amirkabir University of Technology. He did his BSc in Computer Software Engineering, in 2002 at Islamic Azad University of Meibood. He is top manager of VOIP and Firewall Projects.

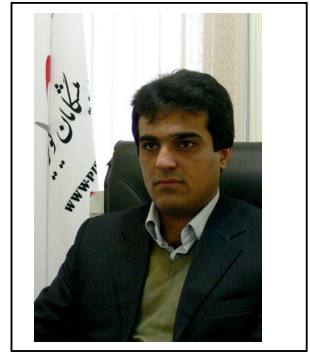

Mohammad Reza Rezaeinezhad is currently a Managing Director of Pishgaman Kavir Yazd Co. he completed his degree in Information Technology Engineering, in 2011 at Shiraz University. He did his BSc in Electronic Engineering, in 1998 at Ferdosi University of Mashhad. He is Chairman of the Board of Pishgaman Kavir Yazd (P.K.Y) cooperative company. It was established in 1996; with goals of reaching top level of Information Technology (IT) .This Company is servicing all internet and communication connections in all provinces in Iran.

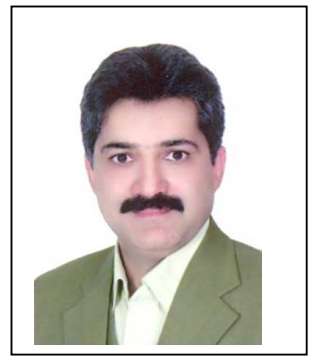